\documentclass[aps,manuscript,showpacs,showkeys,superscriptaddress,nofootinbib]{revtex4-1}
\usepackage{graphicx}
\usepackage{epsfig}  
\usepackage{epsf}    
\usepackage{dcolumn}
\usepackage{bm}
\usepackage{dcolumn}
\usepackage{textcomp}
\usepackage[tbtags]{amsmath}
\usepackage{amsfonts}
\usepackage{float}
\usepackage{subfig}

\usepackage[]{hyperref}
  \hypersetup{
  bookmarks=true,         
  unicode=true,         
  pdftoolbar=true,     
  pdfmenubar=true,     
  pdffitwindow=true,    
  pdfstartview={FitH},    
  pdfsubject={Sterilenus@DUNE},  
  pdfnewwindow=true,     
  pdfcreator={RevTeX},
  colorlinks=true,     
  linkcolor=red,   
  citecolor=blue,    
  filecolor=black,  
  urlcolor=blue,           
  }
\usepackage{hypcap}

\newcommand{\beq}{\begin{equation}}
\newcommand{\eeq}{\end{equation}}
\newcommand{\beqa}{\begin{eqnarray}}
\newcommand{\eeqa}{\end{eqnarray}}

\newcommand{\ty}{{\theta_{13}}}
\newcommand{\tz}{{\theta_{23}}}

\newcommand{\dcp}{\delta_{\mathrm{cp}}}
\newcommand{\ph}{{\phi_{21}}}

\newcommand{\s}{\sigma}

\begin{document}
\title{Octant of $\theta_{23}$ at long baseline neutrino experiments in the light of Non Unitary Leptonic mixing  }
\author{Debajyoti Dutta}
\email[Email Address: ]{debajyotidutta@hri.res.in}
\affiliation{Harish-Chandra Research Institute, Chhatnag Road, Jhunsi, Allahabad 211019, India}
\affiliation{Homi Bhabha National Institute, Training School Complex,
Anushaktinagar, Mumbai - 400094, India}

\author{Pomita Ghoshal}
\email[Email Address: ]{pomita.ghoshal@gmail.com }
\affiliation{Department of Physics, LNM Institute of Information Technology (LNMIIT),\\
Rupa-ki-Nangal, post-Sumel, via-Jamdoli, Jaipur-302 031, Rajasthan, India}

\author{Sandeep K. Sehrawat$^{1, 2,}$}
\email[Email Address: ]{sandeepsehrawat@hri.res.in}
\date{\today}

\begin{abstract}
The octant of the atmospheric mixing angle $\tz$ is still undetermined by neutrino oscillation experiments.
Long-baseline experiments like NO$\nu$A, T2K and DUNE offer good prospects for determining the octant of $\tz$. However, their capability
to do so may be compromised by the possible presence of non-unitarity in the leptonic mixing matrix. In this paper, we study
in detail the octant degeneracy with and without non-unitarity at the level of oscillation probabilities and events for these experiments. We also analyze their octant sensitivity and discovery reach and show that they are hampered in the presence of non-unitarity.

\end{abstract}

\pacs{14.60.St,14.60.Pq,14.60.Lm,13.15.+g}
\keywords{Octant Resolution, Long-Baseline experiments, DUNE, NO$\nu$A, T2K}
\maketitle

\section{Introduction}
\label{Introduction}
Over the past two decades, there have been dramatic advances in the neutrino sector and the neutrino oscillation experiments are constantly focusing on the precise determination of a) the neutrino mass-squared differences, $\Delta m^{2}_{ij} = m^{2}_{i} - m^{2}_{j}$,  and b) the mixing angles $\theta_{ij}$ along with the attendant CP phase $\delta_{cp}$, which constitute the PMNS\footnote{See \cite{Maki, Pont, Gri}} mixing matrix $U_{\alpha i}$, with i, j = 1, 2, 3 where, $i
\neq j$ and $\alpha = e,\mu,\tau$. Although significant success has been achieved towards the precise measurements of the mixing angles, the CP phase $\delta_{cp}$ is yet almost completely unknown.  Of the three mixing angles, $\theta_{12}$, $\theta_{13}$ and $\theta_{23}$ governing the oscillation amplitudes, $\theta_{12}$ and $\theta_{13}$ are measured precisely. But the confusion related to the atmospheric mixing angle $\tz$ is still present and recent results from the NO$\nu$A collaboration have intensified the doubt regarding the non-maximality of $\tz$ \cite{nova1}. If $\tz$ is non-maximal, then $\tz$ is either $< 45^0$ or $> 45^0$. Octant degeneracy \cite{fugli} is the incapability of an experiment to distinguish between $\tz$ and $(\pi/2 - \tz)$ as the $\rm P_{\mu\mu}$ neutrino oscillation probability has the same value for $\tz$ and $(\pi/2 - \tz)$. This is known as the intrinsic octant degeneracy. The leading order terms in this channel are dependent on $\sin^2 2\tz$. The disappearance data through its sensitivity to $\sin^2 2\tz$ can provide stringent constraints. Hence this channel is considered the best for testing the maximality of $\tz$. The octant sensitivity in LBL experiments dominantly comes from the $\nu_e$ appearance channel. Here, the leading order terms depend on the combination of $\sin^22\ty \sin^2\tz$. With the precise measurements of $\ty$ \cite{db, dc, reno}, it is possible to get enhanced octant sensitivity from this channel \cite{hu, mina, sup, animesh, bora}. So the combination of both these channels helps in resolving the octant degeneracy. Octant sensitivity has been studied explicitly in the context of different experiments \cite{raj1, raj2, mono, sanjib1, sanjib2}.

There are several issues which could be indicative of underlying new physics (For a discussion see \cite{Altarelli}). The discovery of neutrino mass itself points to the need of physics beyond the Standard Model. If such new physics is real, then it could affect the interpretation as well as the measured sensitivities of these long baseline experiments. Non-unitarity (NU) in the leptonic mixing matrix is one of the possible departures from the three-neutrino scenario. It can occur due to the induction of neutrino mass through the type-I seesaw mechanism. With such fermionic singlets, the SM neutrinos mix and the full leptonic mixing matrix gets enlarged. But if the new fermionc singlet is heavier than the energy of the experiment, then only the light states propagate and in such a scenario, the $3 \times 3$ leptonic mixing matrix, which is a sub-matrix of the full unitary leptonic mixing matrix, is no more unitary \cite{ Antusch, Fern, Goswami, Forero, Fischer}. In the presence of such a heavy neutral lepton, the sensitivity of these experiments will get affected. So it is important to revisit the capabilities of the long baseline experiments like NO$\nu$A \cite{Ayres}, T2K \cite{Abe}, DUNE \cite{int, Adams}, LBNO \cite{Agarwalla} and T2HK \cite{Abe1} etc in the presence of such new physics. The possible presence of new physics can give rise to new CP phases and these phases mimic the leptonic CP phase. This may lead to further degeneracies. In \cite{Valle, Valle1}, the degeneracy was presented at the probability level, and in \cite{Ge}, the authors had provided the solution to resolve this degeneracy. Recently, CP violation as well as the mass hierarchy sensitivity have been studied in the presence of NU \cite{me1, me2} in long baseline experiments. Some direct and indirect methods to test the unitarity of the PMNS matrix was studied in \cite{Zhang}. Octant degeneracy in the presence of a light sterile neutrino has been studied in \cite{sanjib3}. The extra neutrino considered in \cite{sanjib3} is light enough to be produced in the neutrino beam and hence the octant degeneracy is studied considering the $4 \times 4$ unitary leptonic mixing matrix.
 
 In the present work, we focus on the effect of NU on the measurements of the octant of $\tz$ at long-baseline
experiments - NO$\nu$A, T2K and DUNE. We probe the octant sensitivity of DUNE and NO$\nu$A + T2K in the presence of such a heavy neutral lepton (HNL) called `Sterile' or `right-handed' neutrino. This work is different from \cite{sanjib3} in the sense that the extra neutrinos are too heavy to be produced in the neutrino beam, and hence the effective $3 \times 3$ mixing matrix will be non-unitary and we study the octant sensitivity under the assumption of a non-unitary leptonic mixing matrix. The most general structure of the parametrization adopted in this work starts with \cite{Valle2} and the  symmetrical parametrization technique can be found in \cite{Valle3}. With the help of bi-event plots, we study the octant degeneracy in the presence of NU. We depict how the octant sensitivity changes with baseline in presence of NU. 

The paper is organized as follows: in Section II we discuss the oscillation probabilities $P (\nu_{\mu} \rightarrow \nu_{e})$ and $P (\nu_{\mu} \rightarrow \nu_{\mu})$ relevant to the experiments, explaining the parametrization adopted. We also outline the constraints as well as the true values of the simulation parameters in this section. In Section III, we discuss the octant degeneracy in the presence of NU with the help of bi-event plots. In section IV, we present our octant sensitivity results in terms of sensitivity and discovery plots. We conclude with a discussion of the results in Section V. 
 
\section{Theory}
 In the presence of a Neutral Heavy Lepton, the $3\times 3$ neutino mixing matrix is no longer unitary and gets modified as  \begin{equation}N = N^{NP}U\end{equation} where U is the $3\times 3$ PMNS matrix.  $N^{NP}$ is a left triangular matrix and can be written as \cite{Forero}
 
 \begin{equation}
N^{NP} = 
\begin{pmatrix}
\alpha_{11} & 0 & 0 \\
\alpha_{21} & \alpha_{22} & 0 \\
\alpha_{31} & \alpha_{32} & \alpha_{33} \\
\end{pmatrix}
\end{equation}

Due to this structure of the pre factor matrix, there remain only four extra parameters which affect the neutrino oscillations - the real parameters $\alpha_{11}$ and $\alpha_{22}$, one complex parameter $|\alpha_{21}|$ and the phase associated with $|\alpha_{21}|$. In the presence of the non unitary mixing matrix, the electron neutrino appearance probability changes in vacuum, as explained in \cite{Forero}. $P_{\mu e}$ in terms of these new parameters can be written as (neglecting cubic products of $\alpha_{21}$, $\sin\theta_{13}$,
and $\Delta m^2_{21}$)
\begin{eqnarray}P_{\mu e} = (\alpha_{11}\alpha_{22})^2 P^{3\times 3}_{\mu e}+\alpha_{11}^2\alpha_{22}|\alpha_{21}|P^{I}_{\mu e}+\alpha_{11}^2|\alpha_{21}|^2
\label{eq:Pmue}
\end{eqnarray}
where $P^{3\times 3}_{\mu e}$ is the standard three flavor neutrino oscillation probability and $P^{I}_{\mu e}$ is the oscillation probability containing the extra phase that appears due to the non unitary nature of the mixing matrix. $P^{I}_{\mu e}$ in the above expression can be written as :
 
\begin{eqnarray}
 \begin{split}
 P^{I}_{\mu e} = -2\left[\sin(2\theta_{13})\:\sin\theta_{23}\:\sin\left(\frac{\bigtriangleup m^2_{31}L}{4E_{\nu}}\right)\:\sin\left(\frac{\bigtriangleup m^2_{31}L}{4E_{\nu}}+\phi_{NP}-I_{123}\right)\right]\\
 -\cos\theta_{13}\:\cos\theta_{23}\:\sin(2\theta_{12})\:\sin\left(\frac{\bigtriangleup m^2_{21}L}{2E_{\nu}}\right)\sin({\phi_{NP}})
  \end{split}
  \label{eq:Pmue2}
 \end{eqnarray}
  where, \\
 $I_{123} = -\delta_{cp}$ and $\phi_{NP}$ is the new phase associated with $\alpha_{21}$ such that $\phi_{NP}=\phi_{21} = Arg(\alpha_{21})$.
 
 Again the survival probability $P_{\mu\mu}$ in terms of these NU parameters can be written as:
\begin{eqnarray}
P_{\mu\mu} = \alpha_{22}^4 P^{3\times3}_{\mu\mu} 
          + \alpha_{22}^3|\alpha_{21}|   P^{I_1}_{\mu\mu} 
          + 2|\alpha_{21}|^2\alpha_{22}^2 P^{I_2}_{\mu\mu} 
\label{eq:Pmumu3}
\end{eqnarray}
Where $P^{3\times3}_{\mu\mu} $ is the the standard oscillation formula in vacuum.
The extra terms $ P^{I_1}_{\mu\mu}$ and $ P^{I_2}_{\mu\mu}$ in the oscillation probability are given by:

\begin{eqnarray}
 \begin{split}
 P_{\mu \mu}^{I_1}\approx &-8 \left[
 \sin \theta_{13} \sin{\theta_{23}}\cos(2 \theta_{23})
\cos(\text{I}_{123}-\phi_{NP})\right]\:\sin^2 
{\left(\frac{\Delta m^2_{31}L}{4E} \right)}\\
&+2\left[
\cos \theta_{23} \sin (2 \theta_{12}) \sin^2\theta_{23} \cos(\phi_{NP})\right]
\,\sin{\left(\frac{\Delta m^2_{31}L}{2E} \right)}
\,\sin{\left(\frac{\Delta m^2_{21}L}{2E} \right)} \, ,
 \end{split}
 \label{eq:Pmumu33}
\end{eqnarray}

\begin{equation}
P_{\mu \mu}^{I_2} \approx 1-2\sin^2\theta_{23}\,\sin^2 
{\left(\frac{\Delta m^2_{31}L}{4E} \right)} .
\end{equation}

From Eq.~(\ref{eq:Pmue}) and (\ref{eq:Pmumu3}), it is observed that in the presence of a heavy fermionic singlet, both appearance and disappearance measurements are heavily dependent on the new parameters. In particular, the interplay between the new phase associated with $ |\alpha_{21}|$ and the standard $\dcp$ for some non zero values of $|\alpha_{21}|$ affects the octant sensitivity in these experiments. 

The intrinsic octant degeneracy, in which the probability is a function of $\sin^22\tz$, is defined as \begin{equation}
P(\theta^{tr}_{23}) = P(\pi/2 -\theta^{tr}_{23} ) 
\end{equation}

If there is a $\sin^2\tz$ or $\cos^2\tz$ in the oscillation probability, 
\begin{equation}
P(\theta^{tr}_{23}) \neq P(\pi/2 -\theta^{tr}_{23} )  
\end{equation}

However, this inequality in probabilities can change to an equality and lead to a persisting octant degeneracy for different values of $\ty$ and $\dcp$ in the standard 3$\nu$ case, i.e.

\begin{equation}
P(\theta^{tr}_{23}, \ty, \dcp) = P(\pi/2 -\theta^{tr}_{23}, \theta^{'}_{13}, \dcp^{'} )
\end{equation}
 
Here, $\theta^{tr}_{23}$ is the true value of the mixing angle and all other primed and unprimed parameters lie within their 3$\sigma$ allowed ranges. In the presence of NU, as seen from Eq.~(\ref{eq:Pmue2}) and (\ref{eq:Pmumu33}), there are sine and cosine terms where both the phases $\dcp$ and $\phi_{NP}$ appear together. So these interference terms change the overall probability appreciably. Since, in the presence of NU, the oscillation probability is not only a function of $\dcp$ but also the new phase, we can write the above degeneracy equation as:
\begin{equation}
P(\theta^{tr}_{23}, \ty, \dcp, \phi_{NP}) = P(\theta^{\rm wrong}_{23}, \theta^{'}_{13}, \dcp^{'}, \phi^{'}_{NP} )
\end{equation}
 Where $\theta^{\rm wrong}_{23}$ is the value of the atmospheric mixing angle in the opposite octant. The above expression depicts that the octant degeneracy is further complicated by the presence of the NU phase and its interplay with the standard CP phase in the oscillation probability.

\subsection{Simulation Details}
This analysis is carried out with realistic simulations of all the three experiments using GLoBES \cite{Huber, Kopp},
which includes matter effects as well as relevant systematic uncertainties for each experiment. We have incorporated MonteCUBES's \cite{Blennow} NU Engine (NUE) with GLoBES to perform this analysis.
In this analysis, we consider the best fit values of the oscillation parameters from \cite{Garcia}. The true values of the solar and reactor mixing angles are fixed at $\theta_{12} = 33.48^0$ and $\theta_{13} = 8.5^0$ respectively. 
For assumed NH (IH) as the true hierarchy, the two mass square differences are $\Delta{m}^2_{21} = 7.5 \times 10^{-5}$ e$V^2$ and $\Delta{m}^2_{31} = 2.457 \times 10^{-3}$ e$V^2$ ($-2.457 \times 10^{-3}$ e$V^2$) respectively. The $3\sigma$ allowed range of $\theta_{23}$ is [38.3, 53.3] with a best fit value of $42.3^\circ$ $(49.5^\circ)$ assuming NH (IH) as the true hierarchy.  But we consider one benchmark value of $\tz$ in the $3\sigma$ allowed range such that $\tz = 40.3^\circ$ $ (49.7^\circ)$ in the LO (HO) while generating the bi-event and the discovery plots. Also in the standard 3$\nu$ case, we marginalise over $\dcp$ and $\ty$ in their 3$\s$ allowed ranges. \\

The constraints related to the non unitarity parameters can be found in \cite{Forero, Fischer, enrique1, enrique2}. Strong bounds on the diagonal NU parameters can be obtained from universality constraints which in turn give additional restrictions on the off diagonal NU parameters \cite{enrique1, enrique2}. We note that such bounds are derived considering the charged current induced processes with the assumption that there is no new physics other than the non-unitarity mixing coming from type I see-saw. In the presence of other new physics e.g. right-handed interactions or neutrino-scalar Yukawa interactions in type II see-saw, these bounds are no longer valid. On the other-hand, the neutrino experiments namely NOMAD \cite{nomad} and CHORUS \cite{chorus1, chorus2}, provide direct restrictions on the off diagonal NU parameters which, while less stringent, are model-independent. In this study, we do not pursue the approach which would use the tighter, but model-dependent bounds in \cite{enrique1, enrique2}, but instead, use the model-independent constraints from neutrino experiments \cite{Forero}.

 The bounds that we use in this work are: $\alpha_{11}^2 \geq 0.989$, $\alpha_{22}^2 \geq 0.999$ and $|\alpha_{21}|^2 \leq 0.0007$ at 90\% C.L. \cite{Forero}. The allowed range of $\phi_{NP}$ is $[-\pi, \pi]$. We assume these limiting values as the true values of the NU parameters while generating the bi-probability as well as the bi-event plots. While performing the $\chi^2$ analysis, we have assumed the central values of these allowed ranges as the true values. Since the effect of the third row elements of $\rm N^{NP}$ matrix is negligible even in the presence of matter effect, we have not considered them in our analysis.

In this work, we have studied the octant degeneracy as well as the sensitivity of three long baseline experiments- T2K (Tokai to Kamioka), NO$\nu$A (The NuMI{\footnote{Neutrinos at the Main Injector} } Off-axis $\nu_{e}$ Appearance experiment) and DUNE (Deep Underground Neutrino Experiment) in the presence of NU. The main goal of T2K is to observe
$\nu_{\mu}\rightarrow\nu_{e}$ oscillations and to measure $\ty$ and leptonic CP violation \cite{t2kcp} while NO$\nu$A can measure the octant of $\tz$, the neutrino mass hierarchy, $\ty$ and leptonic CP violation. DUNE, with its 1300 km baseline, can address all these issues with a higher degree of precision. In our recent work \cite{me1, me2}, we have studied the CP violation sensitivity as well as the mass hierarchy sensitivity of these experiments in the presence of NU. We have specified all the experimental and simulation details in \cite{me1}, and will be using the same information for this work also.

\section{Octant Degeneracy with NU}

\begin{figure}[H] 
\centering
\includegraphics[width=0.49\textwidth]{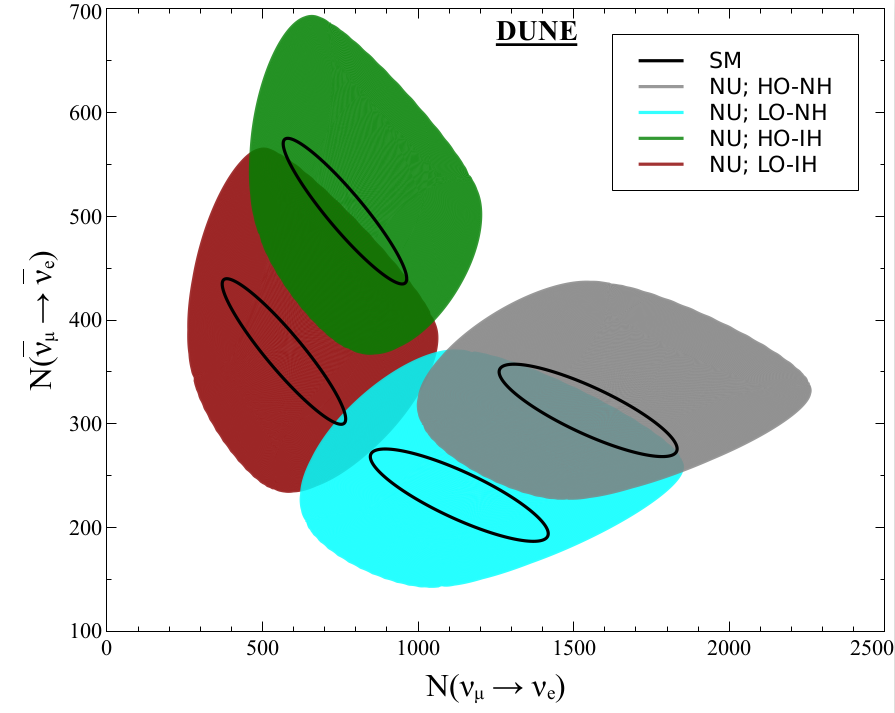}
\includegraphics[width=0.49\textwidth]{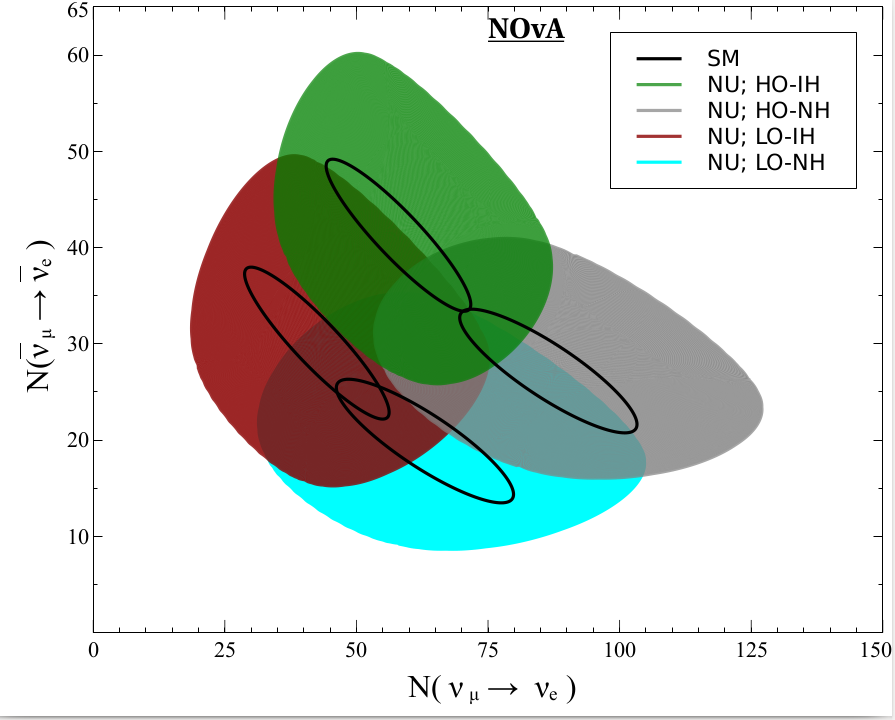}
\includegraphics[width=0.49\textwidth]{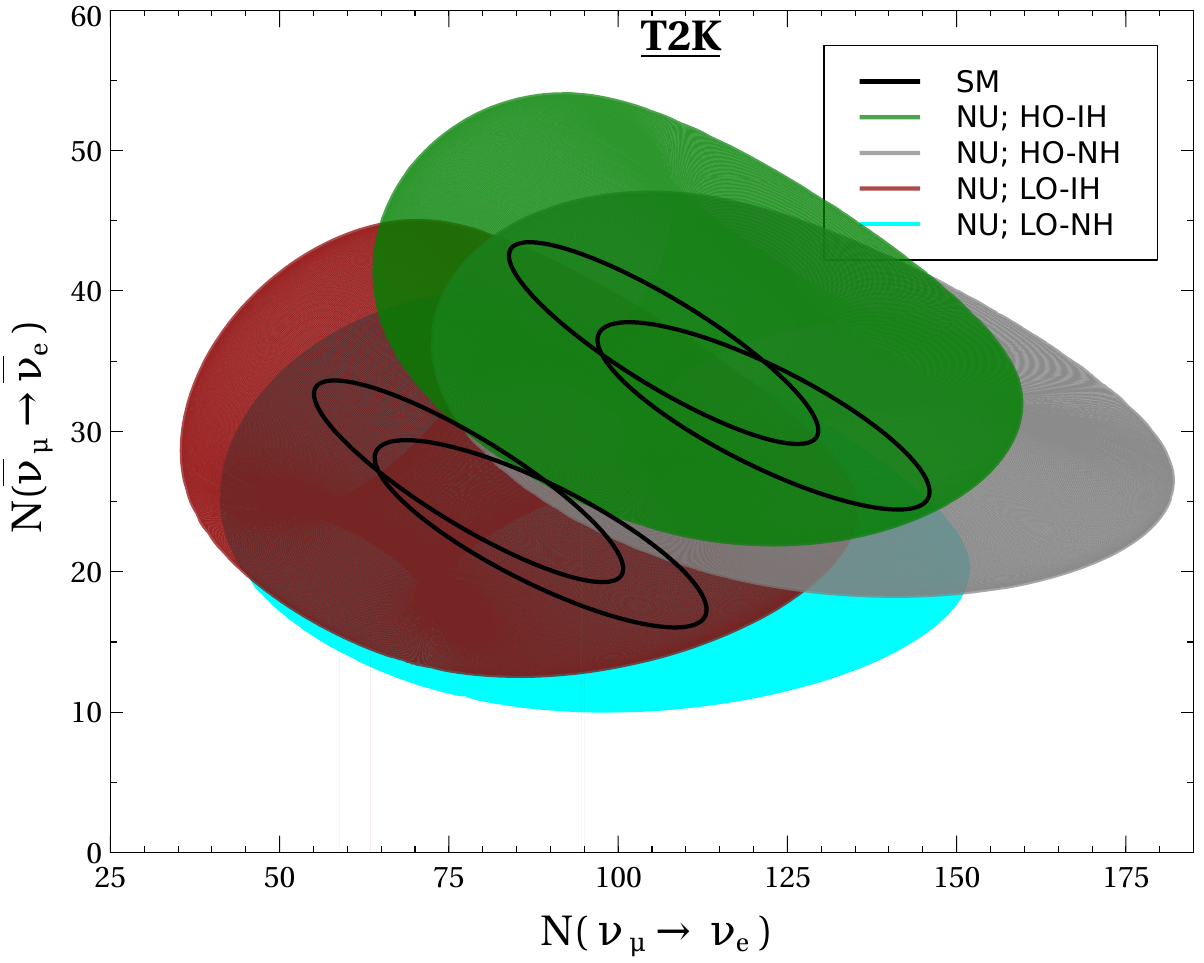}
\caption{\footnotesize{Bi-event plots for DUNE, NO$\nu$A and T2K. The black solid ellipses correspond to the standard 3$\nu$ case and are obtained by varying $\dcp \in [-\pi, \pi]$. The gray (green) and the cyan (darkred) band show the effect of NU for the two cases: HO-NH (HO-IH) and LO-NH (LO-IH). The dotted black ellipses represent the special case when $\phi_{NP}= 0^0$ in both the hierarchies}}
\label{event2}
\end{figure}

For a realistic analysis, we consider the bi-event plots between the energy integrated total number of neutrino and anti-neutrino events (FIG \ref{event2}) for the three experiments. We consider equal $\nu$ and $\bar{\nu}$ runs of 5 years each for DUNE. Both for T2K and NO$\nu$A, we consider 3 years of $\nu$ and 3 years of $\bar{\nu}$ data.

 To generate these plots, we fix $\tz = 40.3 (49.7)$ in the LO (HO) for each assumed hierarchy. Here we have four cases: LO-NH, HO-NH, LO-IH and HO-IH. In FIG. \ref{event2}, four black solid ellipses correspond to these four cases in the three flavor scenario. The gray (green) and the cyan (dark red) bands show the effect of NU for the two cases: HO-NH (HO-IH) and LO-NH (LO-IH). The bands are obtained when both the phases $\dcp$ and $\phi_{NP}$ are varied from $-\pi$ to $\pi$ and all other NU parameters are fixed to their boundary values i.e. $\alpha_{11} = 0.9945$, $\alpha_{22} = 0.9995 $ and $|\alpha_{21}| = 0.0257$. The black dotted ellipses correspond to the special case when $\phi_{NP}=0^0$. For each $\delta_{cp}$, due to the variation of $\phi_{NP}$ in [$-\pi$, $\pi$], we get the continuous band of ellipses. The following observations can be made from FIG. \ref{event2}:
 
 \begin{itemize}
\item  In DUNE, all the four standard ellipses are well separated. It indicates that  DUNE can resolve octant degeneracy in the 3$\nu$ framework. But in the presence of NU the situation changes completely. For assumed $\tz = 40.3^0$ $(49.7)$, there is a significant degeneracy between the two hierarchies in the LO (HO). If we consider the co-ordinate $(1300, 300)$ in the total events plot for DUNE, it lies in the overlapping region between the gray (HO-NH) and the cyan (LO-NH) band, and hence it is not possible to determine whether the co-ordinate belongs to the cyan band or the gray band. Similarly, the co-ordinate (600, 500) is commonly shared by both the green (HO-IH) and the dark red (LO-IH) band. Hence in both the hierarchies, we cannot pin point the exact octant in the overlapping region. NO$\nu$A and T2K also show similar behaviour. The standard 3$\nu$ ellipses are well separated  for the two octants like the bi-probability plots. But, with the new physics scenario, overlapping is more significant.

\item In all the three experiments, it is observed that almost 50$\%$ of the standard ellipse corresponding to HO is a part of the NU induced band of ellipses in the LO. i.e. the co-ordinate (1500, 300) in the case of DUNE is not only a part of the standard HO-NH ellipse, but also a part of both the gray and cyan bands. Similarly the point (700, 500) is common to the HO-IH standard ellipse as well as to both the green and the dark red bands. Hence in presence of NU, a) there is a degeneracy not only between the standard HO and the NU induced HO in a particular hierarchy, but also between the standard HO and the NU induced LO in the same hierarchy.  Similar behaviour can be observed for T2K and NO$\nu$A also. But in DUNE, this is not true in the case of standard LO which is degenerate with the NU induced LO only. In T2K and NO$\nu$A, there is a slight degeneracy between the standard LO and the NU induced HO.
 \end{itemize}

 From the above points, we observe that in the presence of NU, all the experiments under consideration suffer from the octant degeneracy irrespective of the hierarchy. In the next section we depict the octant sensitivity of these experiments in the presence of NU.

\section{Sensitivity Studies}
\subsection{Statistical Details and $\chi^2$ Analysis}
The results presented in this section is based on a $\chi^2$ analysis where we have calculated $\Delta\chi^2$ by comparing the predicted spectra for the alternate hypothesis. We define the statistical $\chi^2$ as: 
\begin{equation}
\chi^2 = \sum_{j=1}^{\rm bins} \sum_{k}^{2}\frac{[\rm N^{j,k}_{\rm true} - \rm N^{j,k}_{\rm test}]^2}{\rm N^{j,k}_{\rm true}}
\end{equation}
In the above expression, $\rm N^{j,k}_{\rm true}$ and $\rm N^{j,k}_{\rm test}$ are the events corresponding to data and fit in the $j^{th}$ bin. For neutrinos and anti-neutrinos, $\rm k$ equals to 1 and 2 respectively. Other statistical details can be found in \cite{me1}. 

To study the octant sensitivity, we vary true $\tz$ in the LO (HO) and test the HO (LO) in the `fit'. All other true values of the three flavor oscillation parameters are fixed at their best fit values. For the standard 3$\nu$ case, we marginalise over $\dcp$ and $\ty$ in their 3$\sigma$ allowed ranges. Since the octant measurement is strongly dependent on $\sin^22\ty$, we add priors on $\sin^22\ty$ ($\sigma = 5\%$) and calculate the minimised $\chi^2$ (i.e. $\chi^2_{\rm min}$) for each true $\tz$  and true $\dcp$ in its 3$\sigma$ allowed range. In the non unitary case, in addition to $\dcp$ and $\ty$, we marginalise over all the three new parameters and the new phase in their allowed ranges. We add the prior to the $\chi^2$ and then calculate the minimised $\chi^2$. For each true $\tz$, we show the maximum and the minimum of $\chi^2_{\rm min}$ which is obtained corresponding to a variation of $\dcp$ in the standard case and the variation of $\dcp$ and new phase $\phi_{NP}$ in the NU case, depicting it as a band. In all cases, we also marginalise over the systematic uncertainties.

\subsection{Octant Sensitivity}
\begin{figure}[H] 
\centering
\includegraphics[width=0.49\textwidth]{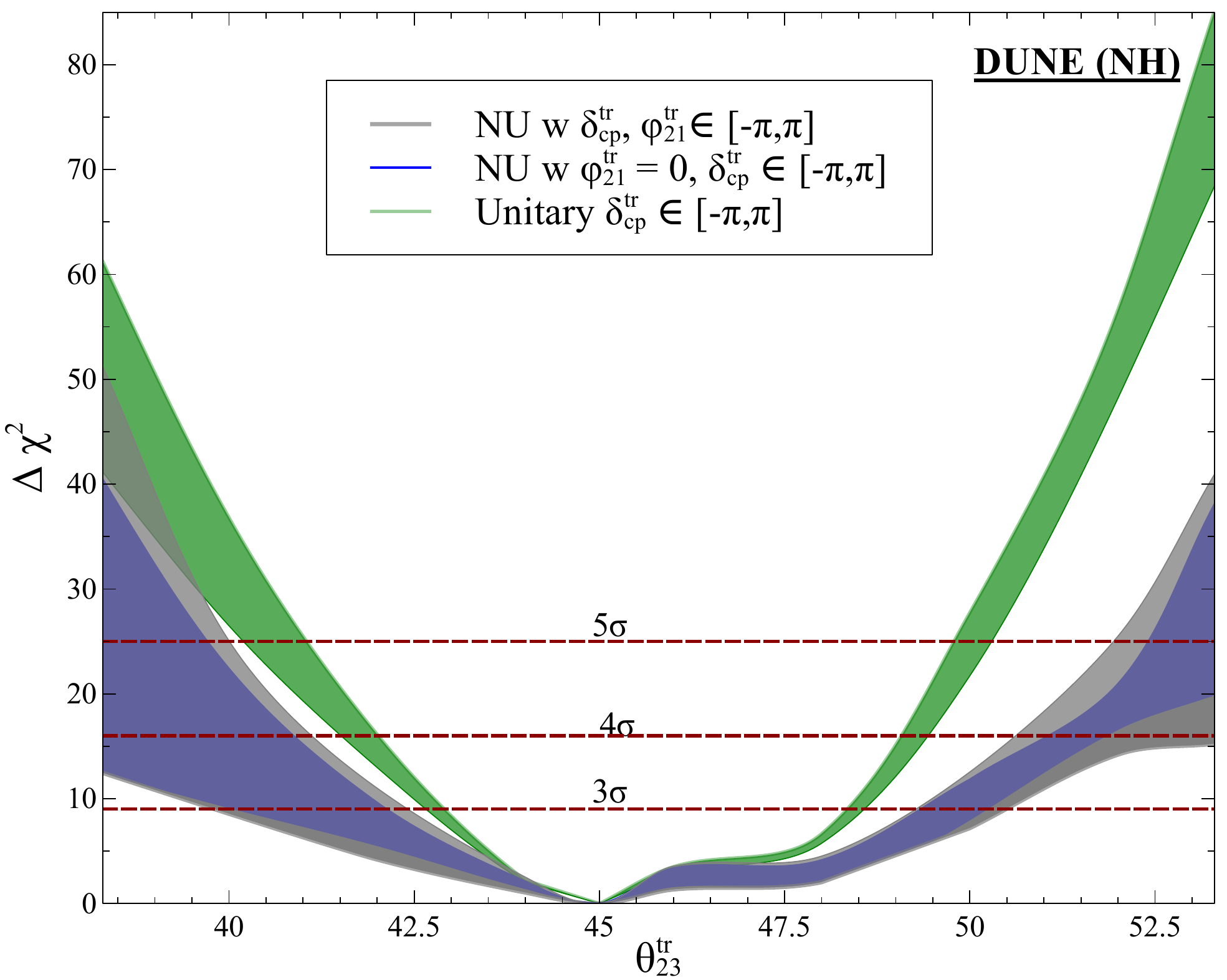}
\includegraphics[width=0.49\textwidth]{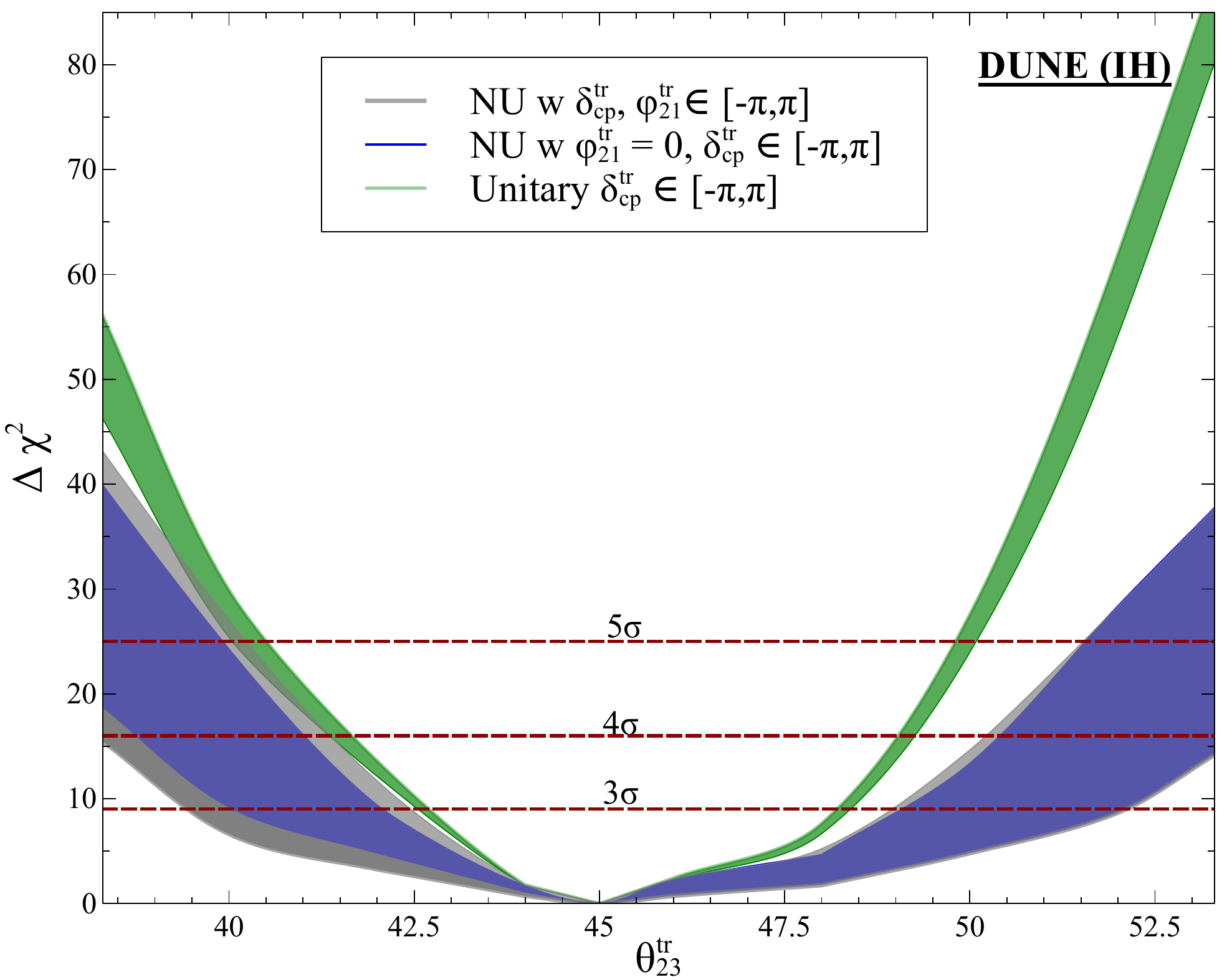}
\includegraphics[width=0.49\textwidth]{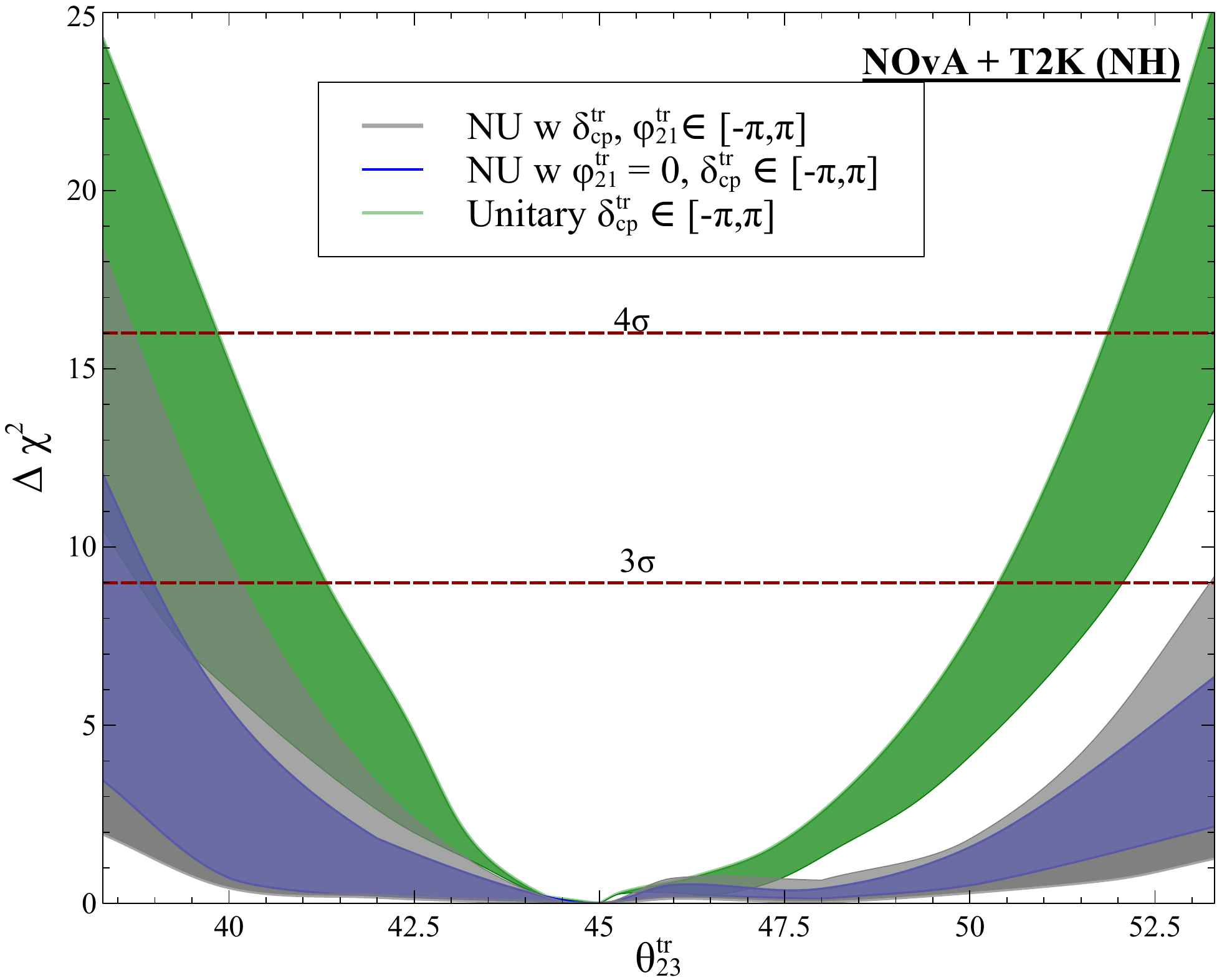}
\includegraphics[width=0.49\textwidth]{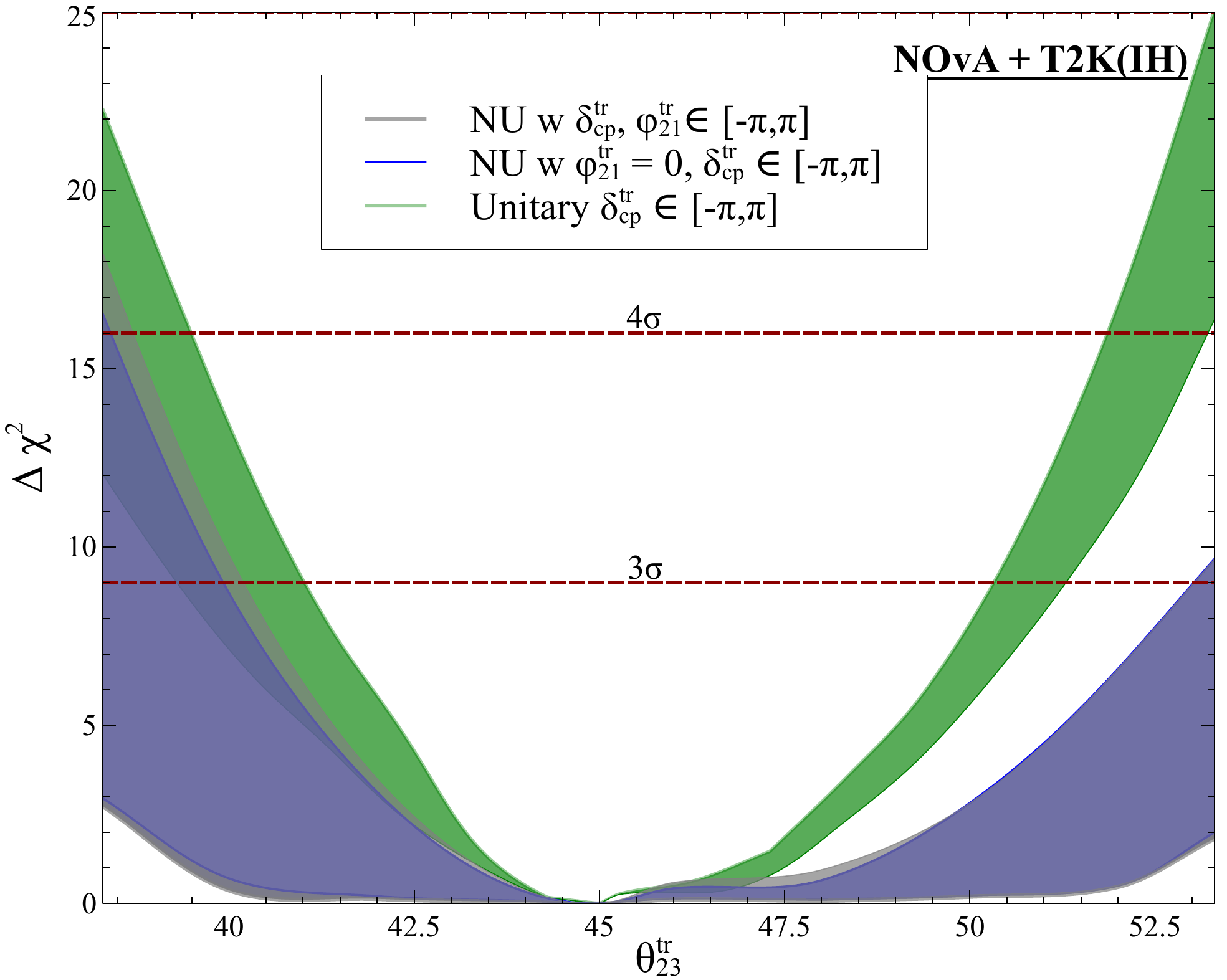}
\caption{\footnotesize{Octant sensitivity plots for DUNE  and NO$\nu$A + T2K for assumed true NH (left) and IH (right). 
We consider the central values of the NU parameters as the true values.}}
\label{sensi1}
\centering
\end{figure}

In this section we present the octant sensitivity results with NU. In FIG. \ref{sensi1} we show the results for DUNE and NO$\nu$A + T2K in both the hierarchies for two cases: a) when all the NU parameters are non zero (gray band), b) when there is no new phase i.e. NU contribution is due to the absolute parameters only (blue band).
 The light green band corresponds to the standard octant sensitivity of these experiments and the width of the band is due to the true variation of $\dcp$ in its full 3$\sigma$ range. The gray band shows the effect of NU and hence the width of the band is due to the variation of both true $\dcp$ and $\phi_{21}$ in $[-\pi, \pi]$ for each true $\tz$. The width of the light blue band is only because of true $\dcp$ in its 3$\sigma$ range as this is a special case where the new phase $\phi_{21}$ is zero in true case.
 To utilize the full potential of these experiments, we have combined both $\nu_{e}$ appearance and $\nu_{\mu}$ disappearance channels both in $\nu$ and $\bar{\nu}$ mode. We make the following observations from these plots:
\begin{itemize}

\item In the upper panel of FIG. \ref{sensi1}, we show the octant sensitivity of DUNE in both the hierarchies. In 3$\nu$ case with true NH (IH), DUNE can resolve octant degeneracy for all true $\dcp$ when $\tz \leq 40.2^0 (40.0^0)$ and $\geq 50.3^0 (50.1^0)$ at 5$\sigma$ C.L.. But in the presence of NU, the octant sensitivity reduces significantly in both the hierarchies. If the leptonic mixing matrix is non unitary, then DUNE cannot resolve the octant ambiguity at 5$\sigma$ or $4\sigma$ C.L. for all values of true $\dcp$. The correct octant can only be identified for all values of $\dcp$ at 3$\sigma$ C.L. when $\tz \leq 39.6^0 (39.4^0)$ and $\geq 50.5^0 (52.3^0)$ for assumed NH (IH).

\item If the new physics phase is zero (only in true case, we still marginalise over $\phi_{21}$ in `fit'), then the octant sensitivity changes slightly and the band corresponding to true $\dcp$ is now a subset of the gray band. But still the sensitivity is less than the standard 3$\nu$ case.

\item In the lower panel of FIG. \ref{sensi1}, we show the octant sensitivity of NO$\nu$A + T2K. We observe that the octant sensitivity in the standard 3$\nu$ case is less compared to DUNE and only for a small fraction of true $\dcp$, they can resolve the octant ambiguity at 4$\s$ C.L.. In the presence of NU, the sensitivity further decreases. In the LO, there is a significant overlapping between the standard and the NU induced case for some true values of $\tz$ in both the hierarchies. But the NU induced sensitivity never exceeds the standard sensitivity.

\item We observe that for some true values of $\tz$ in the LO, there is overlapping between the two scenarios. The overlapping is significant in the case of NO$\nu$A + T2K in both the hierarchies. So in the overlapping region, these experiments cannot tell if the sensitivity is due to a standard value of $\dcp$ or for some combination of $\dcp$ and $\ph$ in the NU framework. Hence the results may be misinterpreted. In the HO, the bands corresponding to standard 3$\nu$ case and the NU case are well separated in both the hierarchies.

\end{itemize}
From the above analysis, it is clear that, due to NU of the leptonic mixing matrix, the capability of the long baseline experiments to resolve octant degeneracy will get hampered. The octant sensitivity decreases as the number of parameters to be marginalised in the `fit' increases with NU. 


\subsection{Octant Discovery}
\begin{figure}[h] 
\centering
\includegraphics[width=0.49\textwidth]{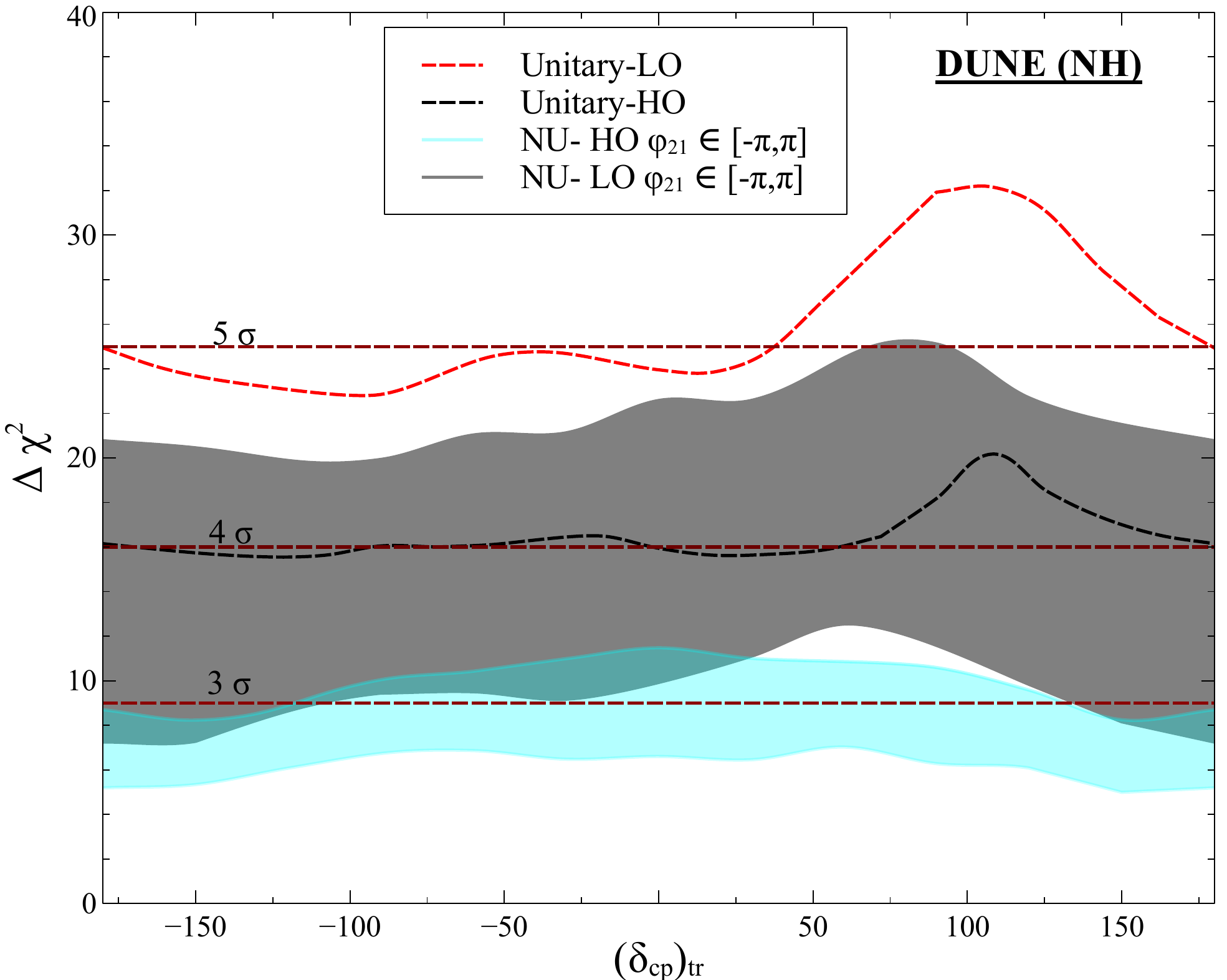}
\includegraphics[width=0.49\textwidth]{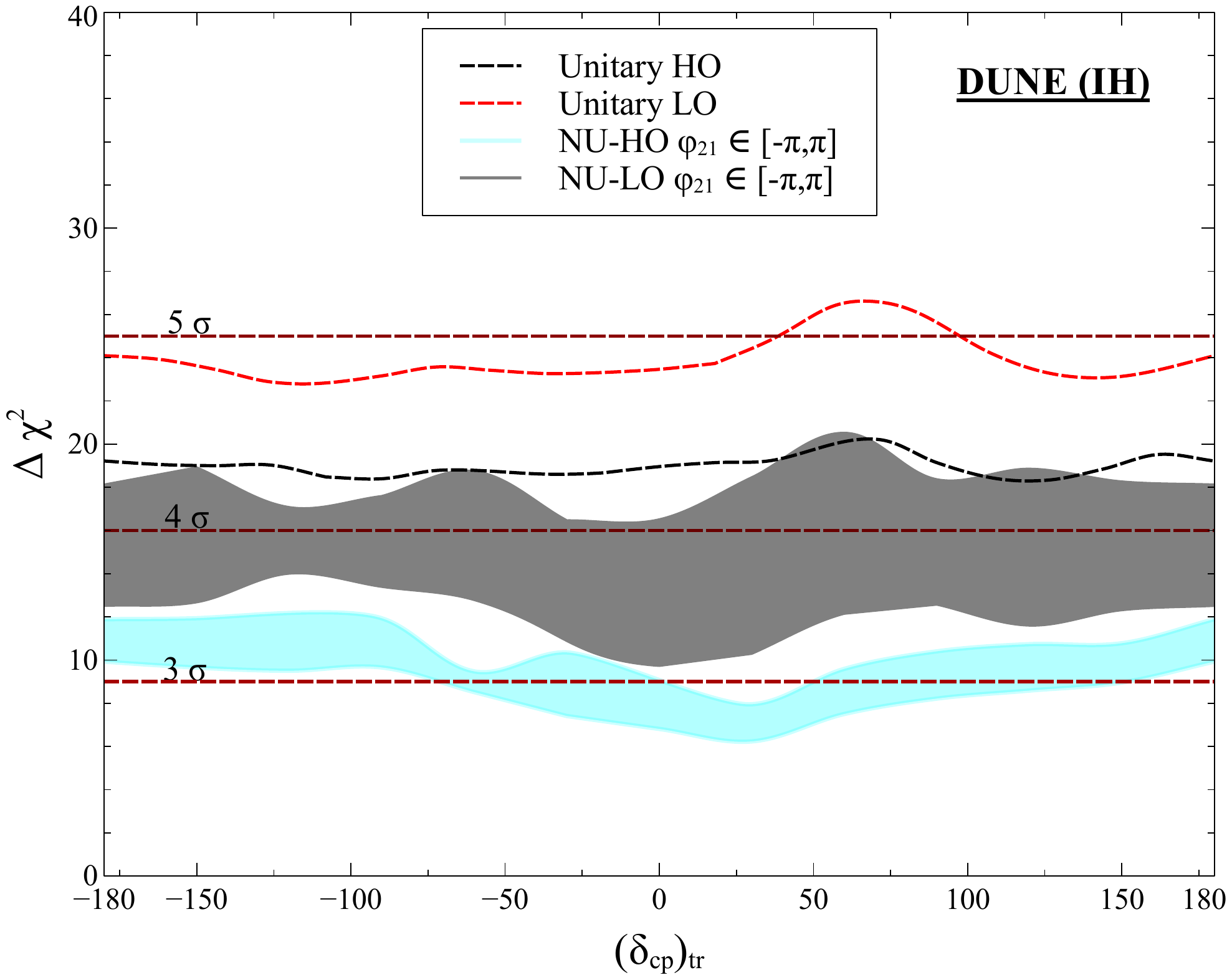}
\includegraphics[width=0.49\textwidth]{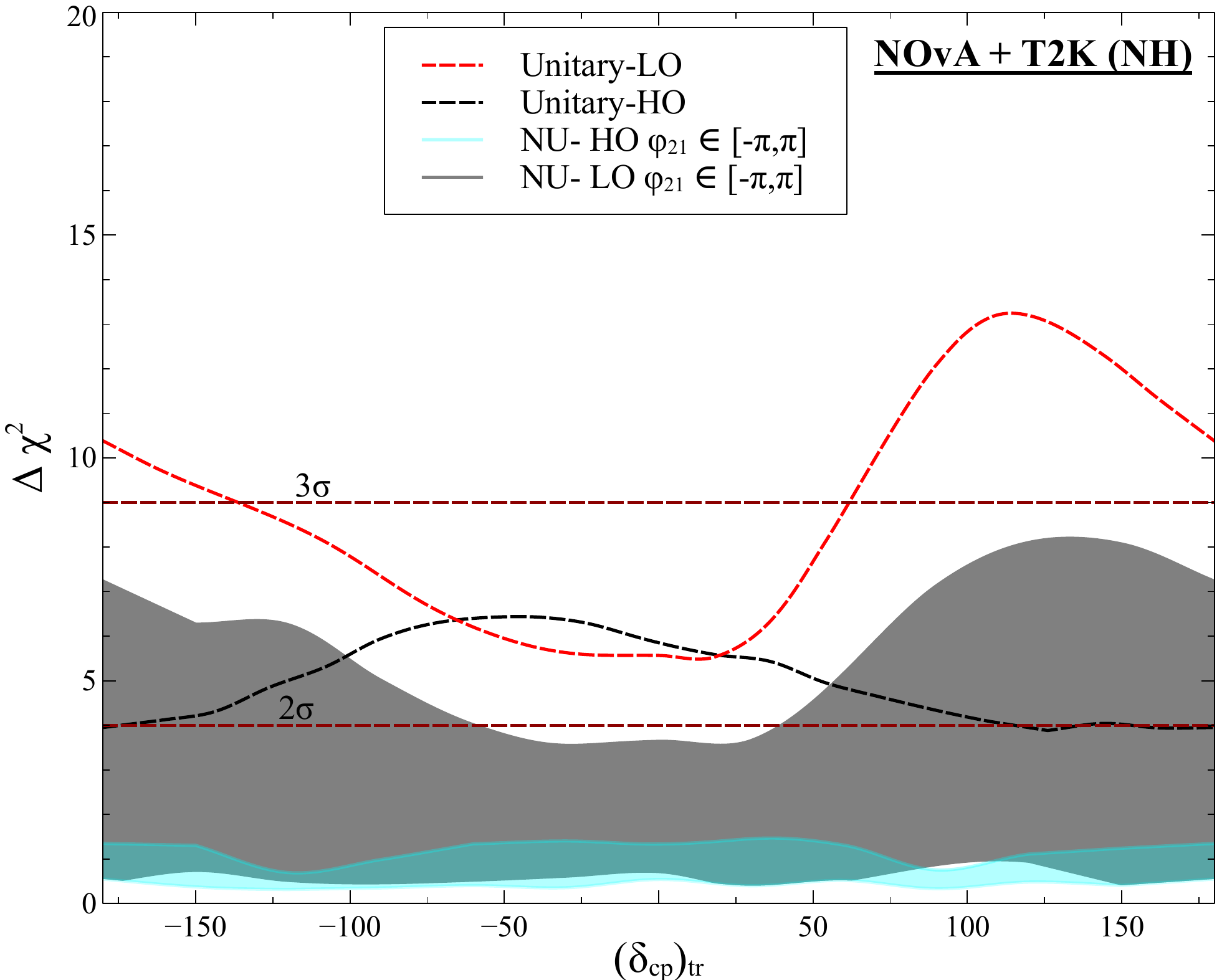}
\includegraphics[width=0.49\textwidth]{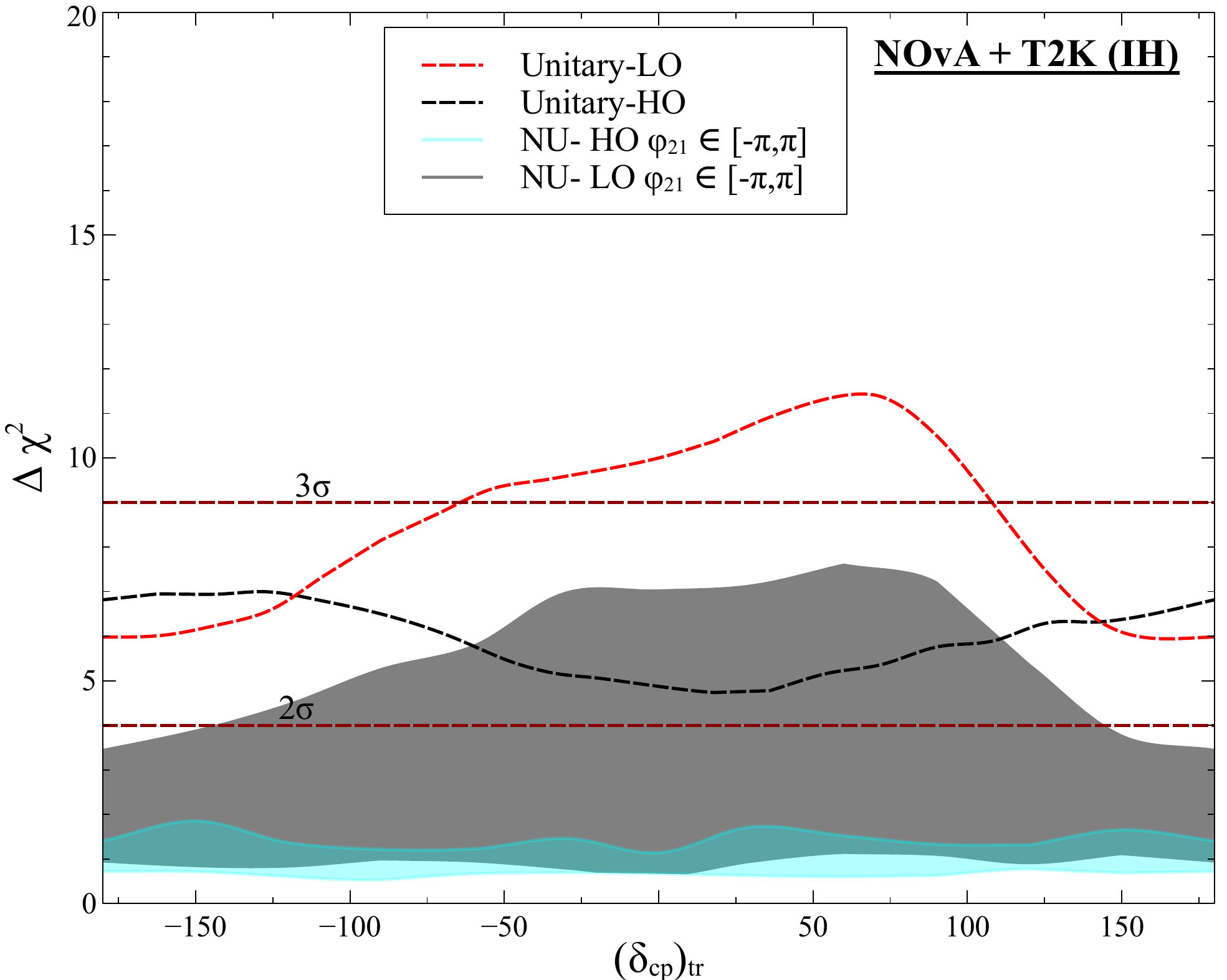}
\caption{\footnotesize{Octant discovery reach of DUNE and NO$\nu$A + T2K for assumed true NH (left) and IH (right). We consider the central values of the NU parameters as the true values.}}
\label{disco1}
\centering
\end{figure}
In this section, we show the octant discovery potential i.e. the potential to exclude the wrong octant at DUNE and NO$\nu$A+T2K. 
We fix the true value of $\tz = 40.3^0$ $(49.7^0)$ in the LO (HO) and vary the test values of $\tz$ in the HO (LO) in the range $[45^0, 53.3^0]$ $([38.3^0, 45^0])$. In the standard case, we marginalise over $\dcp$ and $\ty$ in their allowed 3$\sigma$ ranges and priors are added as explained in the previous section. In the presence of NU, we marginalise over all the new parameters in the `fit' in addition to the standard parameters. Here in the plots, we show the maximum and minimum of the minimised $\chi^2$ for each true $\dcp$, which is obtained by varying true $\ph$ from -$\pi$ to $\pi$. In FIG. \ref{disco1}, we present the octant discovery reach of DUNE and NO$\nu$A+T2K with NU for both the hierarchies. The red (black) dashed line is for true LO (HO) as the true octant in standard framework while the gray (cyan) band represents the same in the NU framework. 

From FIG. \ref{disco1}, we observe that if NH is the established true hierarchy and the LO is the true octant, then DUNE can exclude the wrong octant i.e. the HO with 4$\s$ C.L. for all true $\dcp$ in the standard scenario. But 5$\s$ exclusion is only possible for some fraction of $\dcp$ in the upper half plane (UHP) from $35^0\leq \dcp \leq 180^0$. But in the NU framework, the discovery potential reduces and only for some combinations of true $\dcp$ and $\ph$ a 4$\s$ discovery is possible. Even a $3\s$ discovery of the LO is not possible in the presence of NU for all true $\dcp$. Again in the case of true NH, when the true octant is the HO, DUNE cannot exclude the LO at 4$\s$ C.L. but can discover HO at 3$\s$ C.L. for for all true $\dcp$ in the standard scenario. With NU, discovery potential gets hampered and even a 3$\s$ discovery of the HO is not possible for all combinations of true phases. Again, in the IH mode, the LO discovery potential decreases in DUNE compared to the NH case both in the standard and NU framework. But HO can be discovered at more than 4$\s$ C.L. for all true $\dcp$ in the standard case. 

In the combined case i.e. in NO$\nu$A+T2K, presence of NU decreases the octant discovery potential in both the hierarchies. In standard 3$\nu$ case,  it is almost possible to disfavor the wrong octant for all true $\dcp$ at 2$\sigma$ C.L. for a given octant in both the hierarchies. But, in presence of NU, if LO is the true octant, then only for some combinations of $\dcp$ and $\phi_{21}$, it is possible to disfavor the wrong octant at 2$\sigma$ in both hierarchies. For true HO, situation is more worsen.
\section{Conclusions}
The octant sensitivity of the long-baseline experiments NO$\nu$A, T2K and DUNE
in the presence of NU has been studied in this work. The main conclusions are as follows:
\begin{itemize}
\item If there is a light/heavy sterile neutrino, it will hamper the sensitivity measurements at these long baseline experiments \cite{boris}. Here, we have observed that in the presence of NU, DUNE also undergoes octant degeneracy for both the hierarchies. From the by-event plots it is clear that if the hierarchy is known and leptonic mixing matrix is unitary, then all the three experiments can resolve the octant degeneracy. But, if the leptonic mixing matrix is non unitary, then these experiments cannot resolve the octant degeneracy.  

\item If NU is there, then there will be an additional phase associated with the parameter $|\alpha_{21}|$. As seen from the bi-event plots, due to the degeneracy between the standard ellipse and the NU induced band in a particular octant, it is not possible to determine whether an event corresponding to that octant is a standard 3$\nu$ event or a NU induced event.

\item In the presence of NU, there is more than 50\% probability that the standard measurements of HO in any hierarchy will be affected by the NU induced LO measurements. But the inverse is not true at very long baseline experiments like DUNE. In NO$\nu$A and T2K, there is a slight degeneracy between the standard LO and the NU induced HO.

\item In the NU framework, the octant sensitivity of all the experiments decreases in both the hierarchies. In DUNE, the sensitivity in the presence of NU is good enough to resolve the octant degeneracy with a higher precision, but there is some degeneracy between the standard and the NU framework for some true values of $\tz$. The overlapping is more prominent in case of NO$\nu$A + T2K. In the overlapping region, it is not possible to pinpoint if the sensitivity is due to some standard $\dcp$ or due to some combination of $\dcp$ and $\phi_{21}$. 

\item The octant discovery potential of these experiments also gets hampered in the presence of NU. With NU, DUNE can still manage to exclude the wrong octant with a high degree of precision. But in the case of NO$\nu$A + T2K, the scenario is very poor and they fail to show even a 2$\s$ discovery potential, especially in the case of true HO. 
\item Since recently the error on $\theta_{13}$ has been reduced to  about 3$\%$ by reactor experiments, we have checked our results for the octant sensitivity with a prior of 3$\%$ on $\theta_{13}$. It is found that the effect of changing the $\theta_{13}$ prior from 5$\%$ to 3$\%$  has a minimal effect on the octant sensitivity with NU, but in the standard 3-flavor case the effect is noticeable. This is because in the case of NU, the marginalisation is performed over many parameters in the fit, which minimizes the effect of priors. But in the standard case, since the marginalisation is over $\theta_{13}$ and $\theta_{23}$ only, the effect of the priors on $\theta_{13}$ and $\theta_{23}$ is significant. For this reason, our octant sensitivity results with NU are not significantly affected by the improved precision on $\theta_{13}$.
\end{itemize}
\begin{acknowledgments}
A special thanks to Prof. Raj Gandhi for his useful suggestions and comments in the manuscript. We acknowledge the use of HRI cluster facility to carry out the computations. DD acknowledges the support from the DAE Neutrino project at HRI. DD thanks Suprabh Prakash for some discussions over email. PG acknowledges local support for research at LNMIIT, Jaipur. SKS thanks Mehdi Masud and Samiran Roy for discussions on GLoBES. 
\end{acknowledgments}

\end{document}